\documentclass[a4paper,twoside]{article}

\usepackage{epsfig}
\usepackage{subcaption}
\usepackage{calc}
\usepackage{amssymb}
\usepackage{amstext}
\usepackage{amsmath}
\usepackage{amsthm}
\usepackage{multicol}
\usepackage{pslatex}
\usepackage{url}
\usepackage{apalike}
\usepackage{algorithm2e}
\usepackage[bottom]{footmisc}
\usepackage{SCITEPRESS}     

\begin{document}

\title{Evaluating the Security and Privacy Risk Postures of Virtual Assistants}


\author{\authorname{Borna Kalhor\sup{1}, Sanchari Das\sup{2}}
\affiliation{\sup{1}Department of Computer Engineering, Ferdowsi University of Mashhad, Mashhad, Iran}
\affiliation{\sup{2}Department of Computer Science, University of Denver, Denver, Colorado, USA}
\email{borna.kalhor@mail.um.ac.ir, sanchari.das@du.edu}
}



\abstract{Virtual assistants (VAs) have seen increased use in recent years due to their ease of use for daily tasks. Despite their growing prevalence, their security and privacy implications are still not well understood. To address this gap, we conducted a study to evaluate the security and privacy postures of eight widely used voice assistants: Alexa, Braina, Cortana, Google Assistant, Kalliope, Mycroft, Hound, and Extreme. We used three vulnerability testing tools—AndroBugs, RiskInDroid, and MobSF—to assess the security and privacy of these VAs. Our analysis focused on five areas: code, access control, tracking, binary analysis, and sensitive data confidentiality. The results revealed that these VAs are vulnerable to a range of security threats, including not validating SSL certificates, executing raw SQL queries, and using a weak mode of the AES algorithm. These vulnerabilities could allow malicious actors to gain unauthorized access to users' personal information. This study is a first step toward understanding the risks associated with these technologies and provides a foundation for future research to develop more secure and privacy-respecting VAs.}

\onecolumn \maketitle \normalsize \setcounter{footnote}{0} \vfill
\vspace{-5mm}
\section{\uppercase{Introduction}}
\label{sec:introduction}

Virtual assistants (VAs) are software systems utilizing natural language processing to facilitate human-computer interactions via a series of intents correlated with service interactions and agent utterances~\cite{schmidt2018industrial,guzman2019voices}. Currently, VAs are increasingly integrated into various everyday interactions, such as with smartphones, the Internet of Things (IoT), and smart speakers. Users employ voice commands to perform a spectrum of tasks, ranging from simple activities such as travel planning to more intricate functions such as data analysis and information retrieval~\cite{Modhave_2019}. Despite the growing popularity of VAs, their adoption hinges on the robustness of their security and privacy features, areas that remain underexplored in existing works. Ensuring the privacy and security of VAs is paramount for their continued acceptance and integration into daily tasks by users.

As mentioned above, VAs come with their own security and privacy concerns since they necessitate extensive permissions to execute their designated tasks. For example, they require access to the users' location for navigation, contacts for call initiation, and storage for media playback and display~\cite{burns2019alexa,tan2014poster}. Additionally, they need calendar access for scheduling, background operation permissions for continuous readiness, and network access to interact with various servers~\cite{neupane2022data}. The microphone permission, crucial for VA functionality, introduces a major concern for potentially compromising user privacy~\cite{dunbar2021someone}.

To delve deeper into understanding these vulnerabilities and evaluating the security and privacy postures of VAs, we analyzed eight prevalent voice assistants: Google Assistant (v0.1.474378801), Cortana (v3.3.3.2876), Alexa (v2.2.49), Kalliope (v0.5.2), Mycroft (v1.0.1), Braina (v3.6), Hound (v3.4.0), and Extreme (v1.9.3). Using 
scanners MobSF~\cite{Abraham}, RiskInDroid~\cite{Georgiu}, and AndroBugs~\cite{Lin}, we evaluated their security postures, scrutinized unnecessary permissions, and investigated the destinations of their network communications for potential malicious activity. This analysis aimed to address specific research questions related to these aspects.

\begin{itemize}
  \item Within the contemporary technological landscape, what specific vulnerabilities are prevalent in the architecture and design of leading virtual assistants? Additionally, what preventative strategies can organizations implement to mitigate these vulnerabilities and reduce the risk of malicious exploitation?
  \item As VAs become increasingly embedded in daily digital interactions, how effectively does the application architecture of the most widely utilized VAs safeguard users' sensitive data? Considering existing literature and case studies, what practical measures or adaptations can be suggested and verified to strengthen the intrinsic privacy and security protocols of these assistants?
\end{itemize}

Our study reveals significant vulnerabilities in popular VAs, including improper information management, inadequately secure or absent encryption, lack of certificate validation, unsafe SQL query execution, and insecure communications breaching SSL protocol standards. Based on the findings of this study, we recommend that VA developers implement robust and secure communication and access control mechanisms, in addition to disclosing their privacy practices. Moreover, designers and developers are encouraged to integrate privacy-centric design principles into VAs; this includes privacy-enhancing features, detailed user control over data sharing, and data collection minimization to the essentials of functionality~\cite{osti_10353977}.

In the subsequent sections, we will explore the privacy and security issues of VAs in different aspects, namely code, access control, binary analysis, and tracking. The discovery of weaknesses in this study will be conducive to growing research aimed at fortifying the security and privacy of intelligent AI-powered chatbots and assistants.

\vspace{-5mm}
\section{\uppercase{Related Work}}
\label{sec:relatedwork}
The integration of VAs into various aspects of daily life and work has brought about significant conveniences, but it has also raised substantial concerns about privacy and security. This section reviews research on VA security and privacy, connecting findings from various studies and placing them in a broader context.

Sharif et al. provided a foundational analysis of the privacy vulnerabilities inherent to VAs, identifying six main types of user privacy risks~\cite{sharif2020smart}. They address key concerns such as VAs' constant listening, which may inadvertently gather and store sensitive user data. They also highlight weak authentication mechanisms in VAs, demonstrating how malicious actors can exploit them for unauthorized access to user data or the device. In addition to these user-centric vulnerabilities, the study also draws attention to the potential risks associated with the cloud infrastructure upon which these VAs operate, indicating that it could be susceptible to various forms of cyberattacks~\cite{johnston2014mva,chen2021toward}. Complementing this, the study conducted by Liao et al. takes a closer look at the transparency of VA applications, particularly in terms of how they disclose their privacy policies~\cite{liao2020measuring}. They discovered inconsistencies in disclosure practices, with some applications attempting to access sensitive data without clearly stating so in their privacy policies. 

Lei et al. further expanded the security aspect of VAs, specifically focusing on access control weaknesses in popular home-based VAs such as Amazon Alexa and Google Home~\cite{lei2018insecurity}. Through the execution of experimental attacks, they demonstrated the feasibility of remote adversaries exploiting these vulnerabilities to compromise user security. This study underscores the critical need for stronger authentication mechanisms to safeguard user data and privacy. Shenava et al. extended the discussion to smart IoT devices utilizing virtual assistant technologies, such as Amazon Alexa, to explore potential privacy violations~\cite{shenava2022exploiting}. They created a custom Alexa skill to acquire sensitive user payment data during interactions, illustrating the serious privacy risks associated with custom endpoint functions in VA skills. This research emphasizes the necessity for robust privacy controls and stricter governance over third-party skills to protect user data on smart IoT platforms. 

Adding an empirical dimension to the discussion, Joey et al. conducted a study to understand end-user privacy concerns concerning smart devices~\cite{joy2022investigating}. Their results reveal a substantial gap between user perceptions and the actual privacy practices of these devices, highlighting the importance of improved privacy-preserving techniques and clearer communication from manufacturers. 
Lastly, Cheng et al. provided a literature review focused on the privacy and security challenges associated with Personal Voice Assistants (PVAs), with a special focus on the challenges related to the acoustic channel~\cite{cheng2022personal}. They introduce a taxonomy to organize existing research activities, emphasizing the acute need for enhanced security and privacy protections in PVAs.

Weaving together the findings and insights from these studies, we gain a comprehensive understanding of the multifaceted privacy and security challenges associated with VAs. Constant listening capabilities, weak authentication mechanisms, and transparency issues are identified as prevalent vulnerabilities and issues that need urgent attention. Furthermore, the integration of VAs with smart IoT devices introduces additional complexities and potential avenues for exploitation~\cite{alghamdi2023assessing}. Addressing these challenges requires a concerted effort from researchers, developers, and manufacturers alike, highlighting the need for stronger security protocols, clearer communication regarding data practices, and user education to foster a safer and more secure VA ecosystem. This body of work collectively informs the present study, providing a critical backdrop against which we can further explore and address the security and privacy implications of VAs.

\vspace{-5mm}
\section{\uppercase{Method}}
\label{sec:method}

In our work, we aimed to find security and privacy-centric vulnerabilities prevalent in eight of the most widely used VAs on Android, analyzing their potential exploitation of personal data for purposes such as targeted advertising and user tracking. To select the applications for the study, we employed a selection method that incorporated the number of installations, user ratings, and activity levels in open-source repositories, which resulted in the inclusion of VAs such as Google Assistant, Amazon Alexa, and Microsoft Cortana~\cite{cortana}. We used three tools for the analysis: MobSF, RiskInDroid, and AndroBugs, each providing a unique set of capabilities ranging from malware analysis and API scanning to reverse engineering and application sandboxing.

We specifically selected MobSF, which has been widely used in the previous literature, due to its comprehensive analysis and highly active community on GitHub. AndroBugs, armed with its efficient and robust pattern recognition, enabled us to examine complementary vulnerable areas that were not being examined by MobSF. RiskInDroid was chosen due to its novel approach to using artificial intelligence and its special focus on access control. Although MobSF, AndroBugs, and RiskInDroid each have their strengths, using them in conjunction provided a more thorough vulnerability analysis.

Examining the operational mechanisms of these tools via their GitHub repositories, we discerned that while they all follow a generic approach of decompiling APK files and juxtaposing the source code against a predefined set of rules, MobSF stands out for its more rigorous analysis. MobSF engages in both static and dynamic analysis of applications. In its static analysis phase, MobSF decompiles the APK, transforming it into a pseudocode that mirrors the original source code, facilitating code examination without direct access to the source. This decompiled code is then analyzed to detect potential vulnerabilities using rules drawn from MobSF's active open-source community. Furthermore, MobSF scrutinizes the AndroidManifest.xml file to extract critical information about the application’s components and required permissions, shedding light on potential privacy and security concerns. The tool also evaluates the APIs employed by the application, identifying insecure APIs as potential vulnerabilities. Additionally, MobSF searches for hard-coded sensitive information, such as encryption keys or credentials, which could pose significant security risks. Beyond the decompiled code, MobSF conducts binary analysis to uncover vulnerabilities that might only become apparent during the compilation process~\cite{Appknox_2023}. MobSF also generated a comprehensive report detailing potential vulnerabilities, their severity, and the specific files where these vulnerabilities were detected. 

On the contrary, RiskInDroid provides a risk index for each application, indicating the potential risk to users, with a higher index signifying greater risk. RiskInDroid employs classification techniques using machine learning libraries such as \(scikit-learn\) and classifiers like Support Vector Machines (SVM) and Multinomial Naive Bayes (MNB) to calculate a risk score ranging from 0 to 100. RiskInDroid's unique capability lies in its ability to identify permissions utilized by the application that are not explicitly declared in the manifest files. AndroBugs, on the other hand, does not generate overall assessment metrics for the applications. Analyzing the generated reports, we gained insights into each application's security posture, evaluating its handling of sensitive data, utilization of permissions, and server communications. This enabled us to assess their resilience against potential security threats. Our work underscores the criticality of detecting and mitigating vulnerabilities in VAs as a means to safeguard user privacy and security.

\vspace{-5mm}
\section{\uppercase{Results}}
\label{sec:results}
In our results, we discuss the vulnerabilities we found, focusing on access control, privacy, and database security aspects, among others.

\begin{table*}[t]
\caption{Overview of Critical Security Vulnerabilities in Virtual Assistants.}
\label{tab:vulnerabilities}
\centering
\small
\begin{tabular}{|p{0.6\linewidth}|p{0.3\linewidth}|}
\hline
\textbf{Critical Vulnerability} & \textbf{Virtual Assistants} \\
\hline
Non-SSL URLs & Amazon, Cortana, Extreme, Hound, Kalliope, Mycroft \\
\hline
Implicit Intent Usage & Alexa, Cortana, Extreme, Hound, Mycroft \\
\hline
Missing Stack Canary in Shared Objects & Alexa, Cortana, Hound, Mycroft \\
\hline
Permissive HOSTNAME VERIFIER & Cortana, Extreme, Kalliope \\
\hline
SSL Certificate Validation Bypass & Cortana, Extreme, Kalliope \\
\hline
Raw SQL Queries in SQLite & Cortana, Kalliope \\
\hline
Runtime Command Execution & Cortana, Extreme \\
\hline
WebView Javascript Interface Vulnerability (CVE-2013-4710) & Cortana, Extreme \\
\hline
ContentProvider Exported & Alexa \\
\hline
Base64 Encoding Used as Encryption & Alexa \\
\hline
Misconfigured "intent-filter" & Cortana \\
\hline
ECB Mode AES Usage & Cortana \\
\hline
Clear Text Traffic Enabled & Cortana \\
\hline
Non-Standard Activity Launch Mode & Cortana \\
\hline
CBC Mode AES with Padding Oracle & Extreme \\
\hline
Standhogg 2.0 Vulnerability & Extreme \\
\hline
System-Level Permissions in Manifest & Extreme \\
\hline
Insecure SSL Pinning with Byte Array/Hard-Coded Cert Info & Kalliope \\
\hline
Fragment Vulnerability (CVE-2013-6271) & Mycroft \\
\hline
Debug Mode Enabled & Mycroft \\
\hline
Dangerous Sandbox Permissions & Mycroft \\
\hline
\end{tabular}
\end{table*}

\paragraph{Vulnerabilities in Code} AndroBugs analysis identified several critical vulnerabilities across different VAs. Alexa, for instance, improperly utilizes Base64 encoding as a security mechanism, leading to potential data exposure, as Base64 strings are easily decodable~\cite{lei2017vulnerable}. Furthermore, implicit service checking in Alexa, Cortana, and Hound could allow unauthorized access and execution of sensitive functions, undermining system security. Alexa also exposes its ContentProvider, risking unauthorized data access, and communicates over non-SSL URLs, exposing data to potential DNS hijacking attacks~\cite{shahriar2014content}. Extreme's susceptibility to Strandhogg 2.0 allows malicious apps to impersonate legitimate ones, jeopardizing user security. To mitigate this, setting 'singleTask' or 'singleInstance' launch modes in AndroidManifest.xml is recommended, ensuring single activity instances and thwarting malicious activity duplication.

In our analysis of Cortana and Kalliope, we identified several SSL security vulnerabilities. Both applications permit a user-defined HOSTNAME VERIFIER to indiscriminately accept all Common Names (CN), constituting a critical vulnerability that enables Man in the Middle attacks using a valid certificate~\cite{mitreCWE297Improper}. Furthermore, Cortana initiates connections with 13 URLs that are not secured by SSL. The application also fails to validate SSL certificates properly, accepting self-signed, expired, or mismatched CN certificates. Additionally, the "intent-filter" in the app is misconfigured, lacking associated "actions," thereby rendering the component inoperative~\cite{androidIntentsIntent}.

\begin{table*}[t]
\caption{Trackers utilized by VAs for various purposes. No trackers were found in the remaining VAs.}
\label{tab:trackers}
\centering
\small
\begin{tabular}{|p{4cm}|p{2cm}|p{2cm}|p{2cm}|p{2cm}|}
\hline
\textbf{Tracker Name $\diagdown$ VA name} & \textbf{Alexa}& \textbf{Cortana} & \textbf{Extreme} & \textbf{Hound}\\
\hline
Amazon Analytics (Amazon insights) & Analytics &  & &\\ \hline
Bugsnag & Crash reporting &  &  &\\ \hline
Google Firebase Analytics & Analytics & Analytics & Analytics & Analytics  \\ \hline
Metrics & Analytics &  &  &   \\ \hline
Adjust &  & Analytics &  &  \\ \hline
Google AdMob &  &  & Advertisement &  \\ \hline
Google CrashLytics &  &  & Crash reporting & Crash reporting \\ \hline
Huawei Mobile Services (HMS) Core &  &  & Analytics, Advertisement, Location &  \\ \hline
Facebook Analytics &  &  &  & Analytics \\ \hline
Facebook Login &  &  &  & Identification \\ \hline
Facebook Share &  &  &  & [Unspecified] \\ \hline
Google Analytics &  &  &  & Analytics \\ \hline
Houndify &  &  &  & Identification, Location \\ \hline
Localytics &  &  &  & Analytics, Profiling \\ \hline
OpenTelemetry (OpenCensus, OpenTracing) &  &  &  & Analytics \\ \hline
\end{tabular}
\end{table*}

Our investigation also uncovered a potential for Remote Code Execution due to a critical vulnerability in the WebView "addJavascriptInterface" feature, allowing JavaScript to control the host application. This is particularly hazardous for applications targeting API LEVEL JELLYBEAN (4.2) or lower, as it exposes the application to arbitrary Java code execution with the host application's permissions~\cite{withsecureWebViewAddJavascriptInterface}. In Mycroft, we discovered numerous critical vulnerabilities, including implicit service checking and three URLs that are not secured by SSL. The application is susceptible to the Fragment Vulnerability (CVE-2013-6271), enabling attackers to execute arbitrary code in the application context~\cite{nvd_cve}. Additionally, the "debuggable" property is set to "true" in AndroidManifest.xml, exposing debug messages to attackers via Logcat.

\paragraph{Access Control and Application Permissions}Commonly, VAs require access to sensitive information such as the device's location, storage, and internet connectivity~\cite{alepis2017monkey}. However, it poses a challenge for researchers to ascertain the exact utilization of specific permissions by developers. This ambiguity renders VAs attractive targets for malicious actors, who could exploit these permissions under the guise of legitimate functionality~\cite{bolton2021security}. Identifying an application’s access permissions is relatively straightforward. One can inspect the AndroidManifest.xml file to review the declared permissions necessary for the app’s functionality. Additionally, automated open-source code analysis tools, such as MobSF and RiskInDroid, can reverse engineer and scrutinize the code to identify various permissions. Employing the Android SDK also enables the monitoring of application requests to determine the assets being accessed.

Our study revealed that Extreme utilizes a system-level, special permission, \("WRITE\_SECURE\_SETTINGS"\), as declared in its Android Manifest file. This signature-level permission is typically reserved for applications signed with the system’s signing key, and it grants the app extensive access, allowing it to modify the system's secure settings. Such capabilities present substantial security risks. Generally, special permissions of this nature are granted exclusively to preloaded applications, those integral to the device’s operating system, or apps installed by Google. Furthermore, the \("REQUEST\_INSTALL\_PACKAGES"\) permission observed in Extreme permits the application to request the installation of additional packages. This functionality raises security concerns, as seemingly benign applications might exploit this permission to deceive users into installing malicious packages.

\paragraph{Tracking in VAs}Trackers are software components integrated into applications to collect data on users' activities, capturing details such as the duration of app usage, clicked ads, and more. Often provided by third parties, these trackers can be used to predict future user behavior and interests. Companies utilize this information to tailor their products and services to user needs. Still, trackers can be seen as invasive, raising privacy concerns among users and privacy advocates~\cite{kollnig2021fait}. Nevertheless, developers sometimes employ trackers for benevolent purposes, such as application maintenance. Trackers assist in identifying the most and least effective parts of an application, guiding necessary changes, and are also commonly used for crash reporting~\cite{Farzana_Muhammad_Salman_Zia_Haider_2018}.

Most applications opt not to develop their own tracking systems, instead utilizing established trackers provided by major companies. Our study indicates that a significant number of test subjects employ this technology for a variety of purposes. Google Firebase Analytics emerged as the predominant tracker among the applications studied, primarily utilized for analytics. Google CrashLytics is another prevalent tool used for transmitting crash reports to developers. Additionally, various tracking products developed by Amazon, Google, Facebook, Twitter, Huawei, and others are employed for multiple purposes, including analytics, crash reporting, and user behavior prediction. Table~\ref{tab:trackers} outlines the trackers found in the VAs analyzed in this study and their specific purposes.

\paragraph{Binary Analysis of Shared Libraries}
Each application undergoes numerous repetitive processes and operations beyond its core functionality. Developers often maximize code reusability across different applications by relying on pre-developed code components, such as functions, classes, and variables, collectively referred to as shared libraries. Although this practice offers significant advantages in terms of reusability, it also introduces potential security risks. Malicious actors can exploit vulnerabilities in outdated or improperly maintained libraries to compromise the system. A notable example occurred in November 2021 when security researchers discovered a critical vulnerability in a widely-used Java logging framework, a vulnerability that had remained undetected since 2013—known as the log4shell vulnerability (CVE-2021-44228)~\cite{Wortley_Allison_Thompson_2021}.

\paragraph{Sensitive Data Confidentiality}
In our analysis, we discovered vulnerabilities in several VAs. Braina and Google Assistant are susceptible to CWE-532 and CWE-276, meaning they may inadvertently leak sensitive information into log files or expose such data through other means. Braina additionally poses a security risk due to its read/write operations on external storage, which could be accessed by other applications. Cortana employs the ECB mode for cryptographic encryption, a method considered weak since identical plaintexts yield identical ciphertexts (CWE-327), resulting in predictable ciphertexts. Furthermore, Cortana utilizes the MD5 hashing algorithm, which is known for its susceptibility to collisions. Our findings also indicate that Alexa may expose secrets, such as API keys. Kalliope, on the other hand, is at risk of SQL injection due to its unsafe execution of raw SQL queries, potentially enabling the execution of malicious queries by users. Hound’s domain configuration allows clear text traffic to specific domains, creating a security vulnerability. Google Assistant uses an insecure random number generator (CWE-330) and employs the MD5 algorithm for hashing. Extreme is signed with a v1 signature scheme, rendering it vulnerable to the Janus vulnerability on Android versions 5.0-8.0.
\begin{figure}
\centering
\includegraphics[width=30cm,height=3.8cm,keepaspectratio]{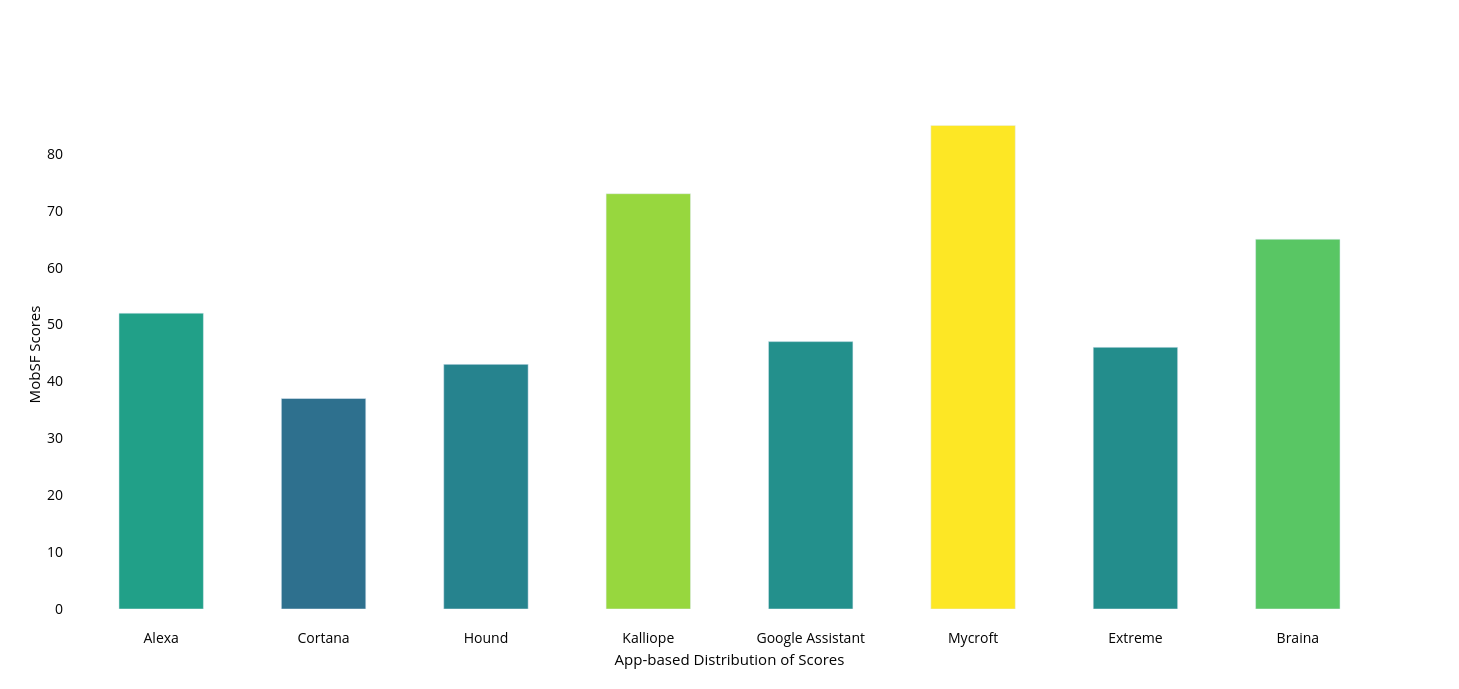}
\caption{App-Based Security Scores Generated by MobSF.}
\label{fig:my_label}
\end{figure}

The comprehensive security scores of the apps, as generated by MobSF, are depicted in Figure~\ref{fig:my_label}. We should note that these scores provide insights into the number and severity of weaknesses identified by the tool only.
\vspace{-5mm}
\section{\uppercase{Discussion}}
\label{sec:discussion-implications}

The pervasive integration of VAs into daily life underscores the crucial need for enhanced security and privacy measures. Our study elucidates a spectrum of vulnerabilities in widely-used VAs, predominantly stemming from insecure communications, misconfigurations, and suboptimal encryption practices. These vulnerabilities not only expose users to potential exploitation but also raise substantial privacy concerns, particularly given the observed instances of user activity tracking and data transmission to third-party services.

This study reveals vulnerabilities in several prevalent VAs, could assist corporations in pinpointing deficient aspects of their software and striving to reinforce them. Additionally, it educates the public on the risks faced by the current status of VAs regarding handling their data and how companies use their data. It also serves as a reminder that the esteemed renown of multinational corporations does not necessarily guarantee the security of their products. Based on this study, we advocate for the adoption of privacy-by-design principles~\cite{sokolova2014privacy} and comprehensive security practices across the software development life cycle (SDLC)\cite{assal2018security}. Specifically, organizations should implement end-to-end encryption for sensitive data, both in transit (using TLS) and at rest (employing robust encryption schemes such as AES-256), and refrain from embedding sensitive information like API keys directly in the code. Regular patching for software and dependencies is paramount, as many exploits target known vulnerabilities for which fixes are readily available.

For developers, embracing secure communication protocols such as HTTPS is imperative to ensure the integrity and confidentiality of data exchanges between VAs and servers~\cite{hadan2019making}. The integration of robust authentication mechanisms, including multi-factor and biometric authentication, adds a critical layer of security, safeguarding user data from unauthorized access~\cite{khattar2019smart,das2018johnny,das2020non}. Transparent and user-friendly privacy policies are vital, enabling users to make informed decisions regarding VA usage. Regular updates to these policies ensure alignment with current data collection and usage practices~\cite{xie2022scrutinizing}. Users, in turn, must be vigilant and proactive in safeguarding their privacy. This includes scrutinizing the permissions requested by VAs, employing strong, unique passwords, and considering multi-factor authentication~\cite{warkentin2011influence,jensen2021multi}. Moreover, explicit user consent should be a prerequisite for the collection of sensitive data. VAs should empower users with greater control over their data, including the ability to delete it, dictate its retention duration, or opt out of data collection entirely~\cite{hutt2023right}. Addressing these challenges necessitates a concerted effort from developers, users, and regulatory bodies aimed at establishing and maintaining a secure and privacy-respecting virtual environment.
\vspace{-5mm}
\section{\uppercase{Limitations \& Future Work}}
\label{sec:limit}
Due to proprietary restrictions imposed by Apple, iOS assistants were not within our investigatory purview. Additionally, there is a limited array of widely used VAs available, generally pre-installed on devices, which inherently narrows the scope of this work. While this study successfully identifies vulnerabilities, we plan to conduct subsequent evaluations to augment these findings through experimental validation. Furthermore, this study predominantly employed static analysis. In our future work, we plan to incorporate dynamic and hardware analysis, utilizing single-board computers to deploy the assistants.
\vspace{-5mm}
\section{\uppercase{Conclusion}}
\label{sec:conclusion}
The proliferation of VAs entails substantial privacy and security ramifications for end-users. This study assessed the security postures and privacy implications of the eight most commonly used VAs, encompassing Alexa, Braina, Cortana, Google Assistant, Kalliope, Mycroft, Hound, and Extreme. Utilizing automated vulnerability scanners, we conducted static code analysis to uncover latent code vulnerabilities and assess the necessity of requested permissions. Our findings reveal that numerous VAs are riddled with critical vulnerabilities, potentially exposing users’ private data to malicious actors. Many VAs exhibit flawed information handling practices and employ suboptimal encryption standards. Our analysis also showed the prevalence of third-party trackers within these applications, highlighting potential data-sharing practices with advertising entities. We underscore the imperative for stringent security measures, advocating for enhanced encryption practices and robust authentication mechanisms. Additionally, we encourage users and developers alike to cognize the inherent privacy risks associated with VA usage, fostering a security-conscious virtual environment.



\vspace{-5mm}
\bibliographystyle{apalike}
{\small
\bibliography{VASecICISSP2024}}

\begin{thebibliography}{}

\bibitem[Abraham, 2023]{Abraham}
Abraham, A. (2023).
\newblock Mobsf/mobile-security-framework-mobsf.
\newblock \url{https://github.com/MobSF/Mobile-Security-Framework-MobSF}.

\bibitem[Alepis and Patsakis, 2017]{alepis2017monkey}
Alepis, E. and Patsakis, C. (2017).
\newblock Monkey says, monkey does: security and privacy on voice assistants.
\newblock {\em IEEE Access}, 5:17841--17851.

\bibitem[Alghamdi and Furnell, 2023]{alghamdi2023assessing}
Alghamdi, S. and Furnell, S. (2023).
\newblock Assessing security and privacy insights for smart home users.
\newblock In {\em ICISSP}, pages 592--599.

\bibitem[Assal and Chiasson, 2018]{assal2018security}
Assal, H. and Chiasson, S. (2018).
\newblock Security in the software development lifecycle.
\newblock In {\em Fourteenth symposium on usable privacy and security (SOUPS 2018)}, pages 281--296.

\bibitem[Bolton et~al., 2021]{bolton2021security}
Bolton, T., Dargahi, T., Belguith, S., Al-Rakhami, M.~S., and Sodhro, A.~H. (2021).
\newblock On the security and privacy challenges of virtual assistants.
\newblock {\em Sensors}, 21(7):2312.

\bibitem[Burns and Igou, 2019]{burns2019alexa}
Burns, M.~B. and Igou, A. (2019).
\newblock “alexa, write an audit opinion”: Adopting intelligent virtual assistants in accounting workplaces.
\newblock {\em Journal of Emerging Technologies in Accounting}, 16(1):81--92.

\bibitem[Chen et~al., 2021]{chen2021toward}
Chen, C., Mrini, K., Charles, K., Lifset, E., Hogarth, M., Moore, A., Weibel, N., and Farcas, E. (2021).
\newblock Toward a unified metadata schema for ecological momentary assessment with voice-first virtual assistants.
\newblock In {\em Proceedings of the 3rd Conference on Conversational User Interfaces}, pages 1--6.

\bibitem[Cheng and Roedig, 2022]{cheng2022personal}
Cheng, P. and Roedig, U. (2022).
\newblock Personal voice assistant security and privacy—a survey.
\newblock {\em Proceedings of the IEEE}, 110(4):476--507.

\bibitem[Das et~al., 2018]{das2018johnny}
Das, S., Dingman, A., and Camp, L.~J. (2018).
\newblock Why johnny doesn’t use two factor a two-phase usability study of the fido u2f security key.
\newblock In {\em Financial Cryptography and Data Security: 22nd International Conference, FC 2018, Nieuwpoort, Cura{\c{c}}ao, February 26--March 2, 2018, Revised Selected Papers 22}, pages 160--179. Springer.

\bibitem[Das et~al., 2020]{das2020non}
Das, S., Kim, A., Jelen, B., Huber, L., and Camp, L.~J. (2020).
\newblock Non-inclusive online security: older adults' experience with two-factor authentication.
\newblock In {\em Proceedings of the 54th Hawaii International Conference on System Sciences}.

\bibitem[Developers, 2023]{androidIntentsIntent}
Developers, A. (2023).
\newblock {I}ntents and intent filters  |  {A}ndroid {D}evelopers --- developer.android.com.
\newblock \url{https://developer.android.com/guide/components/intents-filters}.
\newblock [Accessed 26-10-2023].

\bibitem[Dunbar et~al., 2021]{dunbar2021someone}
Dunbar, J.~C., Bascom, E., Boone, A., and Hiniker, A. (2021).
\newblock Is someone listening? audio-related privacy perceptions and design recommendations from guardians, pragmatists, and cynics.
\newblock {\em Proceedings of the ACM on Interactive, Mobile, Wearable and Ubiquitous Technologies}, 5(3):1--23.

\bibitem[Farzana et~al., 2018]{Farzana_Muhammad_Salman_Zia_Haider_2018}
Farzana, B., Muhammad, D., Salman, A., Zia, U., and Haider, A. (2018).
\newblock Accident detection and smart rescue system using android smartphone with real-time location tracking.
\newblock {\em International Journal of Advanced Computer Science and Applications 9 no}.

\bibitem[Georgiu, 2023]{Georgiu}
Georgiu, G.~C. (2023).
\newblock Claudiugeorgiu/riskindroid.
\newblock \url{https://github.com/ClaudiuGeorgiu/RiskInDroid}.

\bibitem[Guzman, 2019]{guzman2019voices}
Guzman, A.~L. (2019).
\newblock Voices in and of the machine: Source orientation toward mobile virtual assistants.
\newblock {\em Computers in Human Behavior}, 90:343--350.

\bibitem[GVS, 2023]{Appknox_2023}
GVS, C. (2023).
\newblock Binary code analysis vs source code analysis.
\newblock \url{https://www.appknox.com/blog/binary-code-analysis-vs-source-code-analysis}.

\bibitem[Hadan et~al., 2019]{hadan2019making}
Hadan, H., Serrano, N., Das, S., and Camp, L.~J. (2019).
\newblock Making iot worthy of human trust.
\newblock In {\em TPRC47: The 47th Research Conference on Communication, Information and Internet Policy}.

\bibitem[Hutt et~al., 2023]{hutt2023right}
Hutt, S., Das, S., and Baker, R.~S. (2023).
\newblock The right to be forgotten and educational data mining: Challenges and paths forward.
\newblock {\em International Educational Data Mining Society}.

\bibitem[Jensen et~al., 2021]{jensen2021multi}
Jensen, K., Tazi, F., and Das, S. (2021).
\newblock Multi-factor authentication application assessment: Risk assessment of expert-recommended mfa mobile applications.
\newblock {\em Proceeding of the Who Are You}.

\bibitem[Johnston et~al., 2014]{johnston2014mva}
Johnston, M., Chen, J., Ehlen, P., Jung, H., Lieske, J., Reddy, A., Selfridge, E., Stoyanchev, S., Vasilieff, B., and Wilpon, J. (2014).
\newblock Mva: The multimodal virtual assistant.
\newblock In {\em Proceedings of the 15th Annual Meeting of the Special Interest Group on Discourse and Dialogue (SIGDIAL)}, pages 257--259.

\bibitem[Joy et~al., 2022]{joy2022investigating}
Joy, D., Kotevska, O., and Al-Masri, E. (2022).
\newblock Investigating users’ privacy concerns of internet of things (iot) smart devices.
\newblock In {\em 2022 IEEE 4th Eurasia Conference on IOT, Communication and Engineering (ECICE)}, pages 70--76. IEEE.

\bibitem[Khattar et~al., 2019]{khattar2019smart}
Khattar, S., Sachdeva, A., Kumar, R., and Gupta, R. (2019).
\newblock Smart home with virtual assistant using raspberry pi.
\newblock In {\em 2019 9th International Conference on Cloud Computing, Data Science \& Engineering (Confluence)}, pages 576--579. IEEE.

\bibitem[Kollnig et~al., 2021]{kollnig2021fait}
Kollnig, K., Dewitte, P., Van~Kleek, M., Wang, G., Omeiza, D., Webb, H., and Shadbolt, N. (2021).
\newblock A fait accompli? an empirical study into the absence of consent to $\{$Third-Party$\}$ tracking in android apps.
\newblock In {\em Seventeenth Symposium on Usable Privacy and Security (SOUPS 2021)}, pages 181--196.

\bibitem[Labs, 2013]{withsecureWebViewAddJavascriptInterface}
Labs, W. (2013).
\newblock {W}eb{V}iew add{J}avascript{I}nterface {R}emote {C}ode {E}xecution --- labs.withsecure.com.
\newblock \url{https://labs.withsecure.com/publications/webview-addjavascriptinterface-remote-code-execution}.
\newblock [Accessed 26-10-2023].

\bibitem[Lei et~al., 2017]{lei2017vulnerable}
Lei, L., He, Y., Sun, K., Jing, J., Wang, Y., Li, Q., and Weng, J. (2017).
\newblock Vulnerable implicit service: A revisit.
\newblock In {\em Proceedings of the 2017 ACM SIGSAC Conference on Computer and Communications Security}, pages 1051--1063.

\bibitem[Lei et~al., 2018]{lei2018insecurity}
Lei, X., Tu, G.-H., Liu, A.~X., Li, C.-Y., and Xie, T. (2018).
\newblock The insecurity of home digital voice assistants-vulnerabilities, attacks and countermeasures.
\newblock In {\em 2018 IEEE conference on communications and network security (CNS)}, pages 1--9. IEEE.

\bibitem[Liao et~al., 2020]{liao2020measuring}
Liao, S., Wilson, C., Cheng, L., Hu, H., and Deng, H. (2020).
\newblock Measuring the effectiveness of privacy policies for voice assistant applications.
\newblock In {\em Annual Computer Security Applications Conference}, pages 856--869.

\bibitem[Lin, 2023]{Lin}
Lin, Y.-C. (2023).
\newblock Androbugs/androbugs\_framework.
\newblock \url{https://github.com/AndroBugs}.

\bibitem[MITRE, 2023]{mitreCWE297Improper}
MITRE (2023).
\newblock {C}{W}{E} - {C}{W}{E}-297: {I}mproper {V}alidation of {C}ertificate with {H}ost {M}ismatch (4.13) --- cwe.mitre.org.
\newblock \url{https://cwe.mitre.org/data/definitions/297.html}.
\newblock [Accessed 26-10-2023].

\bibitem[Modhave, 2019]{Modhave_2019}
Modhave, S. (2019).
\newblock A survey on virtual personal assistant.
\newblock {\em International Journal for Research in Applied Science and Engineering Technology}, 7(12):305–309.

\bibitem[Neupane et~al., 2022]{neupane2022data}
Neupane, S., Tazi, F., Paudel, U., Baez, F.~V., Adamjee, M., De~Carli, L., Das, S., and Ray, I. (2022).
\newblock On the data privacy, security, and risk postures of iot mobile companion apps.
\newblock In {\em IFIP Annual Conference on Data and Applications Security and Privacy}, pages 162--182. Springer.

\bibitem[NIST, 2013]{nvd_cve}
NIST (2013).
\newblock Nvd - cve-2013-6271.
\newblock \url{https://nvd.nist.gov/vuln/detail/CVE-2013-6271}.
\newblock Accessed on October 27, 2023.

\bibitem[Rahman~Md et~al., 2022]{osti_10353977}
Rahman~Md, R., Miller, E., Hossain, M., and Ali-Gombe, A. (2022).
\newblock Intent-aware permission architecture: A model for rethinking informed consent for android apps [intent-aware permission architecture: A model for rethinking informed consent for android apps].
\newblock {\em ICISSP 2022}.

\bibitem[Schmidt et~al., 2018]{schmidt2018industrial}
Schmidt, B., Borrison, R., Cohen, A., Dix, M., G{\"a}rtler, M., Hollender, M., Kl{\"o}pper, B., Maczey, S., and Siddharthan, S. (2018).
\newblock Industrial virtual assistants: Challenges and opportunities.
\newblock In {\em Proceedings of the 2018 ACM International Joint Conference and 2018 International Symposium on Pervasive and Ubiquitous Computing and Wearable Computers}, pages 794--801.

\bibitem[Shahriar and Haddad, 2014]{shahriar2014content}
Shahriar, H. and Haddad, H.~M. (2014).
\newblock Content provider leakage vulnerability detection in android applications.
\newblock In {\em Proceedings of the 7th International Conference on Security of Information and Networks}, pages 359--366.

\bibitem[Sharif and Tenbergen, 2020]{sharif2020smart}
Sharif, K. and Tenbergen, B. (2020).
\newblock Smart home voice assistants: a literature survey of user privacy and security vulnerabilities.
\newblock {\em Complex Systems Informatics and Modeling Quarterly}, 1(24):15--30.

\bibitem[Shenava et~al., 2022]{shenava2022exploiting}
Shenava, A.~A., Mahmud, S., Kim, J.-H., and Sharma, G. (2022).
\newblock Exploiting security and privacy issues in human-iot interaction through the virtual assistant technology in amazon alexa.
\newblock In {\em International Conference on Intelligent Human Computer Interaction}, pages 386--395. Springer.

\bibitem[Sokolova et~al., 2014]{sokolova2014privacy}
Sokolova, K., Lemercier, M., and Boisseau, J.-B. (2014).
\newblock Privacy by design permission system for mobile applications.
\newblock In {\em The Sixth International Conferences on Pervasive Patterns and Applications, PATTERNS}.

\bibitem[Tan et~al., 2014]{tan2014poster}
Tan, J., Drolia, U., Gandhi, R., and Narasimhan, P. (2014).
\newblock Poster: Towards secure execution of untrusted code for mobile edge-clouds.
\newblock {\em ACM WiSec}.

\bibitem[Warkentin et~al., 2011]{warkentin2011influence}
Warkentin, M., Johnston, A.~C., and Shropshire, J. (2011).
\newblock The influence of the informal social learning environment on information privacy policy compliance efficacy and intention.
\newblock {\em European Journal of Information Systems}, 20(3):267--284.

\bibitem[Warren, 2023]{cortana}
Warren, T. (2023).
\newblock Using cortana on ios or android.
\newblock \url{https://www.theverge.com/2021/4/1/22361687/microsoft-cortana-shut-down-ios-android-mobile-app}.
\newblock Cortana's support on mobile ended on March 31, 2021.

\bibitem[Wortley et~al., 2021]{Wortley_Allison_Thompson_2021}
Wortley, F., Allison, F., and Thompson, C. (2021).
\newblock Log4shell: Rce 0-day exploit found in log4j, a popular java logging package | lunatrace.

\bibitem[Xie et~al., 2022]{xie2022scrutinizing}
Xie, F., Zhang, Y., Yan, C., Li, S., Bu, L., Chen, K., Huang, Z., and Bai, G. (2022).
\newblock Scrutinizing privacy policy compliance of virtual personal assistant apps.
\newblock In {\em 37th IEEE/ACM International Conference on Automated Software Engineering}, pages 1--13.

\end{thebibliography}



\end{document}